\documentclass[prl,twocolumn,preprintnumbers,amsmath,amssymb]{revtex4}

\usepackage{graphicx}
\usepackage{dcolumn}
\usepackage{bm}

\newcommand{\lco}{LaCoO$_3$}

\newcommand{\lscoref}{La$_{0.82}$Sr$_{0.18}$CoO$_3$}

\newcommand{\ch}{susceptibility}

\newcommand{\tad}{thermal expansion}
\newcommand{\Tad}{Thermal expansion}

\newcommand{\tgdown}{$t_{2g}^{\,\,\downarrow}$}
\newcommand{\egup}{$e_{g}^{\,\uparrow}$}
\newcommand{\tg}{$t_{2g}$}
\newcommand{\eg}{$e_{g}$}
\newcommand{\LS}{$t_{2g}^{\,6}e_{g}^0$}
\newcommand{\IS}{$t_{2g}^{\,5}e_{g}^1$}
\newcommand{\HS}{$t_{2g}^{\,4}e_{g}^2$}

\begin{document}

\title{Evidence for a Low-Spin to Intermediate-Spin State Transition in \lco }


\author{C.~Zobel, M.~Kriener, D.~Bruns, J.~Baier, M.~Gr\"{u}ninger, and T.~Lorenz}
\affiliation{ II.~Physikalisches Institut, Universit\"{a}t zu K\"{o}ln, Z\"{u}lpicher Str. 77, 50937 K\"{o}ln, Germany}

\author{P.~Reutler and A.~Revcolevschi}
\affiliation{Laboratoire de Physico-Chimie de l'Etat Solide, Universit\'e Paris-Sud, 91405 Orsay Cedex, France}

\date{\today}

\begin{abstract}

We present measurements of the magnetic \ch\ and of the \tad\ of a
\lco\ single crystal. Both quantities
show a strongly anomalous temperature dependence. Our data are consistently
described in terms of a spin-state transition of the Co$^{3+}$ ions
with increasing temperature from a low-spin ground state (\LS ) to an
intermediate-spin state (\IS) without (100\,K - 500\,K) and with ($>500\,$K) orbital degeneracy.
We attribute the lack of orbital degeneracy up to 500\,K
to (probably local) Jahn-Teller distortions of the CoO$_6$ octahedra.
A strong reduction or disappearance of the Jahn-Teller distortions
seems to arise from the insulator-to-metal transition around 500~K.

\end{abstract}

\pacs{PACS: }

\maketitle

Transition-metal oxides have fascinating physical properties as e.g.
high-temperature superconductivity in the cuprates
or colossal magnetoresistance in the manganites.
Their properties are often governed by a complex interplay of charge,
magnetic, structural, and orbital degrees of freedom.
Moreover, for a given oxidation state some transition metals display
different spin states as it is the case in various cobalt oxides.
Quite recently a class of layered cobalt compounds with the chemical composition
REBaCo$_2$O$_{5+\delta}$ (RE\,=\,rare earth) has
attracted considerable interest. These compounds show a broad variety of
ordering phenomena and other transitions, e.g.
(antiferro- and/or ferro-) magnetic order, charge and/or orbital order,
metal-insulator transitions or spin-state
transitions~\cite{vogt00a,suard00a,moritomo00a,martin97a,respaud01a,wu01a,wang01a,roy02a,akahoshi99a}.
For TlSr$_2$CoO$_5$ it has been proposed
that a metal-insulator transition is driven by a spin
disproportionation, which consists of an
alternating ordering of Co$^{3+}$ ions in an intermediate-spin state
(IS: \IS ; S\,=\,1) and in a high-spin state
(HS: \HS ; S\,=\,2)~\cite{doumerc99a,foerster01a}.

The occurrence of Co$^{3+}$ in different
spin states is known since the 1950s from \lco ~\cite{jonker53a,bhide72a},
which transforms with increasing temperature
from a non-magnetic insulator to a paramagnetic insulator around 100\,K
and shows an insulator-to-metal transition around 500\,K.
But even for this rather simple pseudo-cubic perovskite the nature of these
transitions is still unclear. The ground state is usually
attributed to the low-spin configuration (LS: \LS ; S\,=\,0) and the paramagnetic
behavior above 100\,K to the thermal population of an excited state. However, the
question whether the excited state has to be identified with the HS or the
IS state is subject of controversial discussions.
Early publications often assume a LS/HS
scenario~\cite{asai94a,itoh94a,yamaguchi96a}.
In order to explain the insulating nature up to 500\,K an ordering of
LS and HS Co$^{3+}$ ions has been proposed which vanishes at the
insulator-to-metal transition~\cite{senaris95a,raccah67a}.
Yet the presence of a HS configuration
below 400\,K has been questioned on the basis of X-ray absorption and
photoemission experiments~\cite{abbate94a}.  
Alternative descriptions of \lco\ favoring a LS/IS
scenario~\cite{potze95a,saitoh97b,asai98a,yamaguchi97a,kobayashi00b} are mainly
based on the results of LDA+U calculations~\cite{korotin96a}, which propose
that due to a strong
hybridization between Co-\eg\ levels and O-$2p$ levels the
IS state is lower in energy than the
HS state. Within this scenario the occurrence of orbital order
and its melting have been proposed in order to explain the insulating nature below
500\,K and the insulator-to-metal transition, respectively~\cite{korotin96a}.
Up to now there is no experimental evidence, neither for an orbitally ordered
state, nor for a HS/LS superstructure.

In this paper we present a study of the thermal expansion $\alpha$ and
of the magnetic \ch\ $\chi$ of a \lco\ single crystal. The combined analysis of
$\alpha$ and $\chi$ gives clear evidence for a thermal population of the IS state
{\em without} orbital degeneracy. The lack of orbital
degeneracy could arise from orbital order as proposed in
Ref.~\cite{korotin96a}
or it can be interpreted as a consequence of Jahn-Teller (JT)
distortions of the CoO$_6$ octahedra with Co$^{3+}$ in the JT-active IS state.
Above 600\,K, our $\chi(T)$ analysis suggests
the presence of the IS state {\em with} orbital degeneracy, which may arise from
a suppression or strong reduction of the JT distortion
due to the insulator-to-metal transition.

The crystal used in this study was cut from a large single crystal ($l$\,$\simeq$\,$8$\,cm;
$\o$\,$\simeq$\,$6$\,mm) grown by the floating-zone technique in an image furnace.
The crystal is strongly twinned as usual for distorted perovskites.
The magnetization has been measured by
a SQUID magnetometer in the temperature range from 2\,K up to 300\,K in an applied
field of 50\,mT and by a Faraday balance in the temperature range from 200\,K up to 1000\,K in a
field of 1\,T. A high-resolution measurement of the linear
thermal expansion $\alpha$\,$=$\,$\frac{1}{L} \frac{\partial L}{\partial T}$ has been performed
using a capacitance dilatometer from 4\,K to 180\,K.

\begin{figure}[t]
\begin{center}\includegraphics[width=7.7cm,clip]{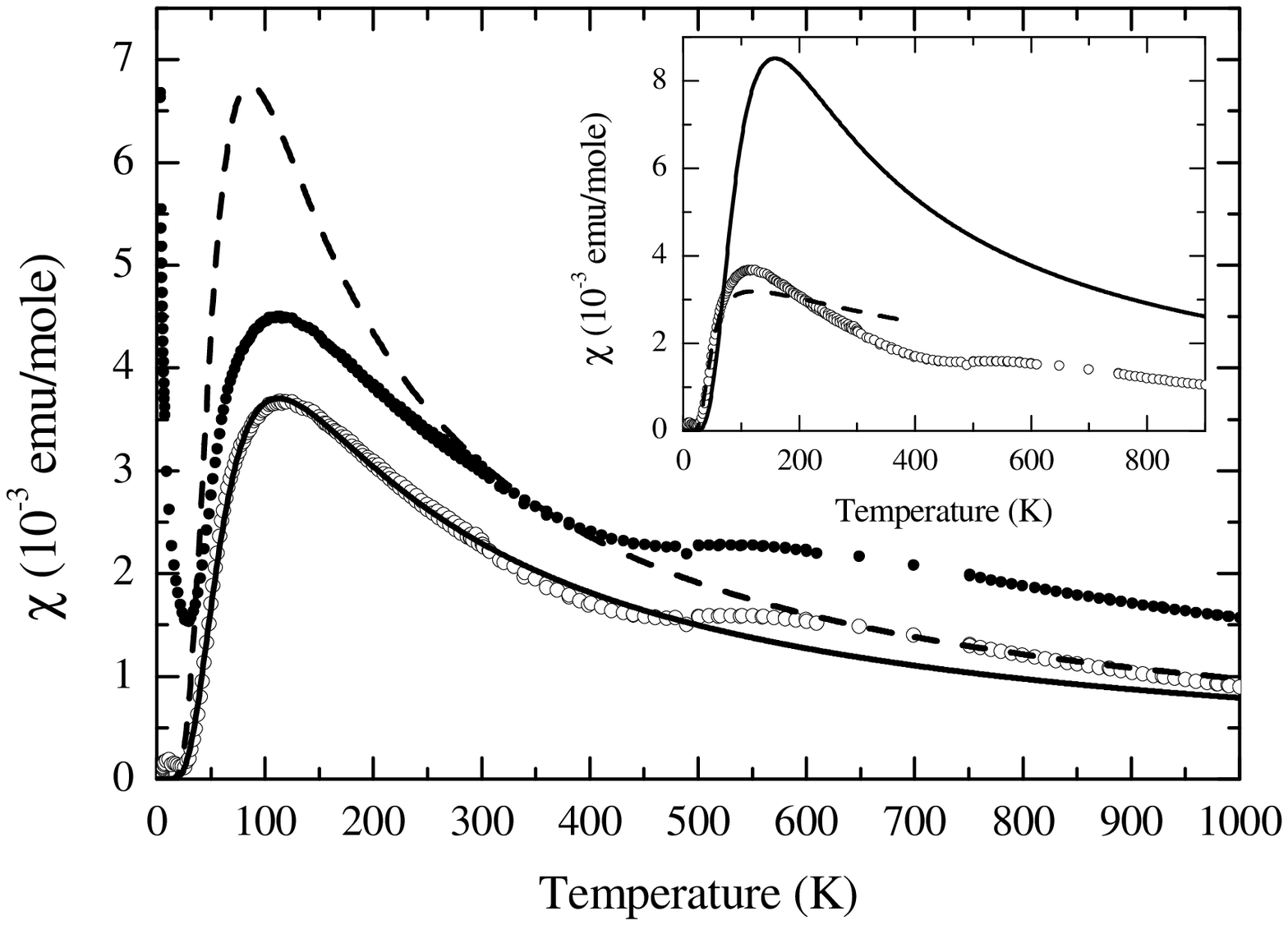}\end{center}
 \caption[]{Total\,($\bullet$) and Curie \ch\ $\chi^C$($\circ$) of \lco .
  The solid line is a fit of $\chi^C(T)$ up to 400~K for a LS/IS scenario
  with $S$\,=\,1, $\Delta$\,$\simeq$\,$180$~K, $g$\,$\simeq$\,$2.1$, and
  $\nu$\,$=$\,$1$ in Eq.~\ref{twolevchi}.
 The dashed line is calculated by increasing the
 orbital degeneracy to $\nu$\,=\,3 and leaving all other parameters fixed.
 The inset shows calculations of $\chi^C(T)$ for a
 LS/HS scenario (see text).
} \label{sus}
\end{figure}

The magnetic susceptibility of our \lco\ crystal (Fig.~\ref{sus}) agrees well
with that found in previous
studies~\cite{bhide72a,itoh94a,yamaguchi96a,saitoh97b}.
The maximum around 100\,K signals the spin-state transition
of the Co$^{3+}$ ions. For higher temperatures $\chi(T)$ shows (i)~a Curie-like decrease
up to about 500\,K, (ii)~a temperature-independent plateau between 500\,K and 600\,K and
(iii)~again a Curie-like decrease above 600\,K. The increase of $\chi(T)$ below 30\,K arises most
probably from a Curie contribution due to magnetic impurities and/or oxygen
non-stoichiometry. In order to obtain the Curie \ch\ $\chi^C$ of Co$^{3+}$
we subtract a term $(P/T+\chi_0)$ from the raw data.  Here
$P/T$ represents the impurity contribution and
$\chi_0=\chi_{\rm dia}+\chi_{\rm vV}$ the sum of the diamagnetic
contribution of the core electrons and of the paramagnetic van Vleck \ch\
of Co$^{3+}$.
A fit of the low-temperature
data gives $P$\,$=$\,$0.02$\,emuK/mole and $\chi_0$\,$=$\,$6.5\cdot 10^{-4}$\,emu/mole.
The magnitude of $P$ allows to estimate an impurity content
of less than 1\%~\cite{polarons}.
Our value of $\chi_0$ is close to those observed e.g. in Ref.~\cite{itoh94a} in \lco\
or in ZnCo$_2$O$_4$ with Co$^{3+}$ in the LS state~\cite{miyatani66}.

The \ch\ of a system with a non-magnetic ground state and a
magnetic excited state reads
\begin{equation}
\chi^C(T) = \frac{N_{\rm A}g^2\mu_{\rm B}^2}{3k_{\rm B}T}\,
 \frac{\nu S(S+1)(2S+1)e^{-\Delta /T} }{1+\nu (2S+1)e^{-\Delta /T} } \; .
\label{twolevchi}
\end{equation}
Here $N_{\rm A}$ is the Avogadro number, $\mu_{\rm B}$ the Bohr magneton and $k_{\rm B}$
the Boltzmann constant, $\Delta $ denotes the energy splitting of the two states,
$g$ is the Land\'{e} factor, $S$ the spin and $\nu$ the orbital
degeneracy of the excited state. For simplicity we consider a purely ionic model
for \lco\ and the spin-only values for the
magnetic moments, where the ground state of Co$^{3+}$ is the LS state
with S\,=\,0 and the electronic configuration \LS . The excited state is
either the IS (S\,=\,1 and \IS ) or the HS (S\,=\,2 and \HS ) state.
In the HS (IS) state the \tgdown\ (\tgdown\ and \egup) levels are only partially
filled. Therefore the HS state consists of 3 orbital states, which are degenerate in
a cubic crystal field leading to $\nu$\,$=$\,$3$. The IS state contains 6 orbital
states but even in a cubic field these states are split into two orbital triplets
separated by an energy of about 1\,eV due to the coulomb interaction within
the 3d shell~\cite{sugano}. Thus, an orbital degeneracy $\nu$\,$=$\,$3$ is expected for the IS
state, too. The orbital degeneracy is lifted by lower symmetries and
the crystal symmetry of \lco\ is only rhombohedral.
It is, however, unclear whether this can be observed in $\chi^C(T)$.
First, the rhombohedral
distortion of \lco\ is very small. Second, the crystal field of the Co$^{3+}$
levels is mainly determined by the surrounding O$^{2-}$ ions and it is not clarified
whether the CoO$_{6}$ octahedra are distorted, i.e. whether the local symmetry
of the Co$^{3+}$ ions is less than cubic.
From neutron scattering only one Co-O distance is reported~\cite{thornton86a}, whereas
optical data give evidence for different Co-O distances~\cite{yamaguchi97a}.
This question is further discussed below. Our
{\em combined} analysis of $\chi$ and $\alpha $ will allow an unambiguous decision which of
the four scenarios, LS/IS or LS/HS with or without orbital degeneracy
yields the appropriate description of the spin-state transition of \lco .

The inset of Fig.~\ref{sus} shows an attempt to describe $\chi^C(T)$
within a LS/HS scenario. Note that Eq.~\ref{twolevchi} has only $\Delta $
as a free parameter (The $g$-factor may be varied only to some extent).
In order to reproduce the strong increase of $\chi^C$
below 100\,K one has to use an energy splitting $\Delta $\,$\simeq $\,$ 290$\,K but in this
case the calculated $\chi^C (T)$ for $T$\,$>$\,$100$\,K is
much larger than the experimental data
(solid line in the inset of
Fig.~\ref{sus}). In this calculation we have set $g$\,$=$\,$2$ and $\nu$\,$=$\,$1$.
Assuming an orbital degeneracy $\nu$\,$=$\,$3$ of the HS state
even increases the discrepancy between experimental and calculated data. There are some
possibilities
to improve the description within a LS/HS scenario. One is to use $g$\,$\simeq$\,$1.1$ (and
$\Delta$\,$ \simeq $\,$ 190$\,K) but such a small $g$-factor is very unlikely~\cite{chargeorder}.
Another one is to introduce an antiferromagnetic nearest-neighbor
coupling $J_{\rm AF}$ as has been done in Ref.~\cite{yamaguchi96a}.
On a mean field level this leads to
\begin{equation}
\chi^{MF}(T) = \frac{\chi^C(T)}{1+ \frac{k_{\rm B}}{N_{\rm A}g^2\mu_{\rm B}^2}
\, z\,J_{\rm AF}\,\chi^C(T)}
\label{MF}
\end{equation}
where $\chi^C(T)$ is given by Eq.~\ref{twolevchi} and $z$ is the number of nearest
neighbors. A fit according to Eq.~\ref{MF} for $T$\,$\le$\,$ 400$\,K is shown by the dashed
line in the inset of Fig.~\ref{sus}. By setting $g$\,$=$\,$2$, $\nu$\,$=$\,$1$ and $z$\,$=$\,$6$ we
obtain $\Delta $\,$ \simeq $\,$ 220$\,K and $J_{\rm AF}$\,$\simeq $\,$56$\,K
(a fit for $\nu$\,$=$\,$3$ yields $\Delta \simeq 270$\,K and $J_{\rm AF}$\,$\simeq $\,$62$\,K).
Such a strong {\em antiferromagnetic} coupling is, however, in clear contradiction to neutron
scattering experiments which give evidence for a weak {\em ferromagnetic}
coupling~\cite{asai94a}. From these arguments we conclude that a LS/HS scenario
does not give a consistent description of the
magnetic properties of \lco\ up to about 500~K.

In contrast to the LS/HS scenario, a good description is found for $\chi^C(T)$
within a LS/IS scenario.
This is shown by the solid line in the main panel of Fig.~\ref{sus}. By setting $\nu=1$
the fit yields
$\Delta \simeq 180$\,K and a reasonable $g\simeq 2.1$. When
an orbital degeneracy $\nu=3$ of the IS state is assumed, the quality of the fit becomes
worse (not shown in Fig.~\ref{sus}). Thus, the fit of $\chi^C(T)$ favors an orbitally
non-degenerate IS state below 500~K. Orbital degeneracy might become
important above 600~K as indicated by the dashed line in Fig.~\ref{sus}, which is obtained by
'switching on' the orbital degeneracy (setting $\nu=3$) and leaving the
other parameters unchanged
($\Delta \simeq 180$\,K and $g\simeq 2.1$). Obviously, the dashed line
is quite close to the experimental $\chi^C(T)$ for $T>600$\,K.

\begin{figure}[t]
\begin{center}\includegraphics[width=7.7cm,clip]{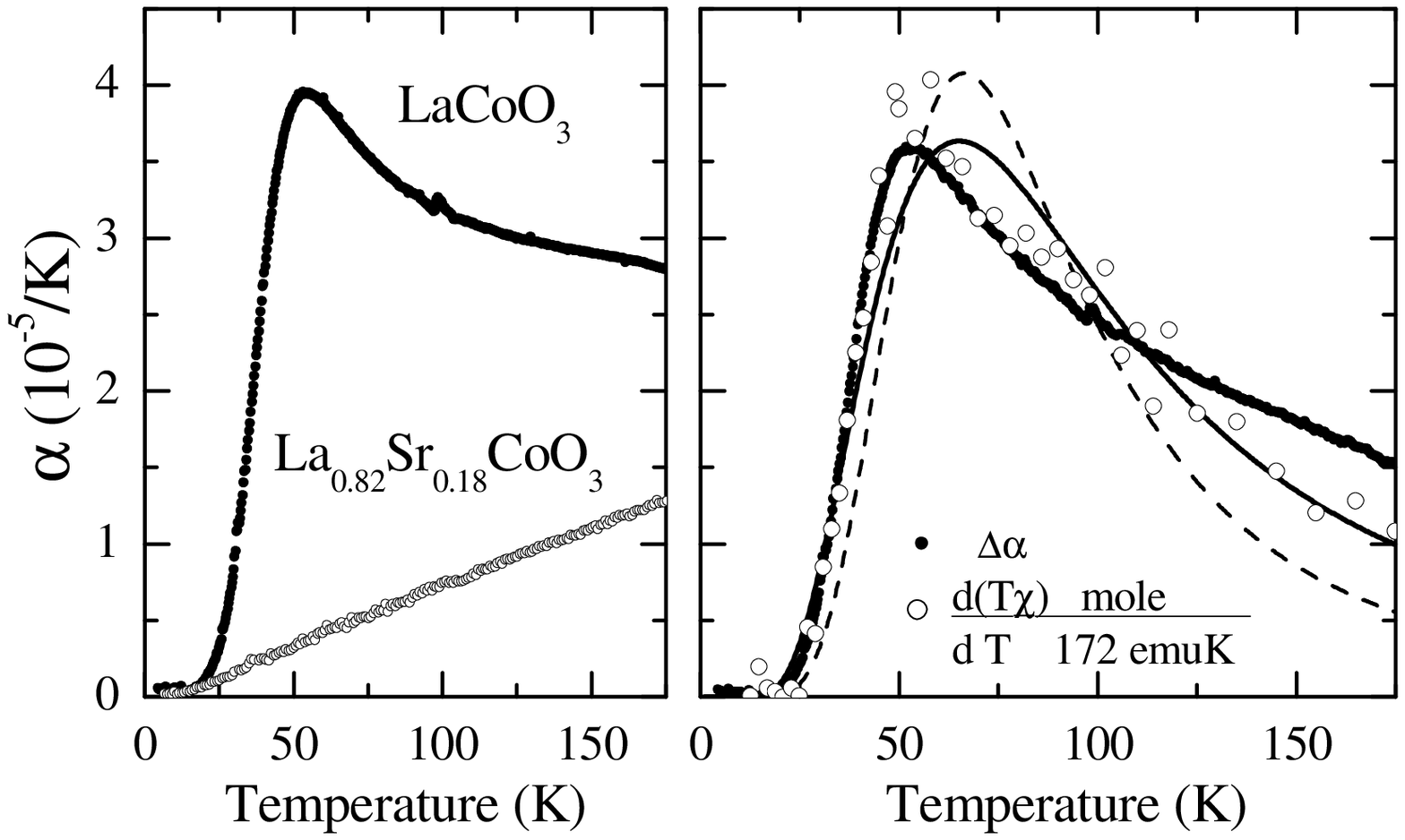}\end{center}
 \caption[]{Left: \Tad\ of pure\,($\bullet$) and of Sr-doped \lco\,($\circ$).
  Right: Anomalous \tad\ $\Delta\alpha$\,($\bullet$) of \lco\ obtained by
  subtracting $\alpha$ of \lscoref . The lines are fits of $\Delta\alpha$ for a LS/IS scenario
  without ($S$\,=\,$\nu$\,=\,1; solid line) and for a LS/HS scenario
  with orbital degeneracy ($S$\,=\,2, $\nu$\,=\,3; dashed line; see Eq.~\ref{twolevalp}).
  The open symbols ($\circ$) show
  $\frac{\partial \, (T \chi^C(T))}{\partial T}$ scaled on $\Delta\alpha$, i.e. there is a scaling
  relation $C \, \Delta\alpha(T)$\,=\,$ \frac{\partial \, (T \chi^C(T))}{\partial T}$ with
  $C$\,=\,$172$\,emuK/mole.} \label{alp}
\end{figure}

In Fig.~\ref{alp} we show the linear \tad\ of \lco\ and of \lscoref . Our
high-resolution data confirm previous results obtained by neutron
diffraction~\cite{asai94a} but allow a more detailed analysis. Whereas $\alpha$
of  \lscoref\ shows a weak monotonous increase with temperature as expected for
ordinary solids the \tad\ of \lco\ is highly unusual: It is rather large and has
a pronounced maximum around 50~K.
In view of the spin-state transition that occurs in \lco\ (but not in  \lscoref )
a straightforward
interpretation of the anomalous behavior of $\alpha$ can be given. In the LS state
of Co$^{3+}$ all electrons
occupy \tg\ levels whereas in the IS and HS state \eg\ levels are also occupied.
Since the \eg\ states are oriented towards the surrounding negative O$^{2-}$
ions a population
of Co$^{3+}$ in the IS (or HS) state causes an additional widening of the lattice.
Simply speaking, the ionic
radius of Co$^{3+}$ depends on its spin state and increases with increasing number of
electrons in the \eg\  states
($r^{\rm HS}_{\rm Co^{3+}} > r^{\rm IS}_{\rm Co^{3+}} > r^{\rm LS}_{\rm Co^{3+}}$).

Note that it is not possible to detect a sharp anomaly of $\alpha$ at a characteristic
temperature in agreement with specific heat data which also do not show such an
anomaly~\cite{stolen97a}. That means, the spin-state transition in \lco\ is not a
phase transition in the thermodynamic sense. This justifies the description
of the spin-state transition by a thermal population
of an excited magnetic state from the LS state which remains the state of
the lowest energy (as assumed in Eq.~\ref{twolevchi}). The (additional) relative length change
due to the spin-state transition is proportional to the thermal population of the
excited state, and the anomalous \tad\ is given by its temperature derivative, i.e. by
\begin{equation}
\Delta \alpha(T) = d \;  \frac{\Delta\;  \nu\, (2S+1)\,e^{-\Delta /T}}{T^2\,(1+\nu\, (2S+1)\,e^{-\Delta /T})^2} \;.
\label{twolevalp}
\end{equation}
The product $\nu (2S+1)$ is the total degeneracy of the excited IS (HS) state and
$d$ is determined by the different Co--O bond lengths for Co$^{3+}$
in the IS (HS) and in the LS state, respectively.

The anomalous \tad\ $\Delta \alpha$ of \lco\ shown in the right panel of
Fig.~\ref{alp} is obtained by subtracting $\alpha$ of \lscoref\ from the raw data.
With respect to $\Delta \alpha$ the different scenarios only differ by the total
degeneracy $\nu(2S+1)$, which amounts to 3 and 5 for the IS and HS state without
and to 9 and 15 for the IS and HS state with orbital degeneracy, respectively.
In Fig.~\ref{alp} we only show the fits for a LS/IS scenario without
($\nu (2S+1)$\,=\,3; solid line) and for
a LS/HS scenario with orbital degeneracy  ($\nu (2S+1)$\,=\,15; dashed line).
The two other fits are lying between the solid and the dashed curve.
The fit for $\nu (2S+1)$\,=\,3 gives the best description of the experimental $\Delta \alpha(T)$
and yields $\Delta$\,$ \simeq $\,$185$\,K in good agreement with
$\Delta $\,$ \simeq $\,$ 180$\,K from the fit of $\chi^C$. The deviations above 50\,K may
arise from the uncertainty in background
determination and/or a temperature dependence of $\Delta $.
Depending on the model the values of $d$ vary between $0.66$\,\% and $0.38$\,\% (see
table~\ref{werte}). They are much smaller than the difference of the Co--O bond lengths of 3\%
obtained from the sums of the tabulated ionic radii~\cite{shannon76} of O$^{2-}$ and Co$^{3+}$
in the LS and the HS state ($1.89$\,\AA\ and $1.95$\,\AA), respectively, giving further
evidence against a LS/HS and for a LS/IS scenario in \lco , because the Co--O bond length
of the IS state is expected to lie between those
of the LS and the HS state of Co$^{3+}$.

The separate fits of both $\chi^C(T)$ and $\Delta\alpha(T)$
already favor a LS/IS scenario without orbital degeneracy but much
more convincingly this conclusion
is obtained by a scaling behavior between both quantities.
From Eqs.\,\ref{twolevchi} and~\ref{twolevalp} a straightforward calculation yields
\begin{equation}
C \, \Delta\alpha(T) = \frac{\partial \, (T \chi^C(T))}{\partial T}
\label{scaling}
\end{equation}
with $C$\,=\,$ \frac{N_{\rm A}g^2\mu_{\rm B}^2}{3k_{\rm B}}
\frac{S(S+1)}{\, d}$. As shown in Fig.~\ref{alp} this scaling
behavior is well fulfilled by the experimental data. The scaling
unambiguously reveals that the anomalous \tad\ of \lco\ arises
from the spin-state transition. Moreover, the scaling factor
$C^{exp}$\,=\,$172$\,emuK/mole agrees almost perfectly with
$C$\,=\,$167$\,emuK/mole expected for a LS/IS scenario without
orbital degeneracy (see table~\ref{werte}). In contrast, the
other three scenarios yield strongly different scaling factors
ranging from 252 to 870\,emuK/mole.  Thus, our data clearly
exclude descriptions within a LS/IS scenario with orbital
degeneracy~\cite{saitoh97b,asai98a} and also within a LS/HS
scenario with or without orbital
degeneracy~\cite{asai94a,itoh94a,yamaguchi96a,senaris95a}.

\begin{table}[t]
\begin{tabular}{c|c|c|c|c}
  &   \multicolumn{2}{c|}{LS / IS: \hskip2mm $S=1$ }  &   \multicolumn{2}{c}{LS / HS: \hskip2mm $S=2$ }   \\
       &  $\nu=1$ & $\nu=3$ & $\nu=1$ & $\nu=3$  \\ \hline
  $d$~(\%)  & 0.66 & 0.44 & 0.55 & 0.38 \\
  $\Delta $~(K)  & 185 & 265 & 205 & 256 \\ \hline
  $C \left(\frac{\rm emuK}{\rm mole}\right)$   & 167 & 252 & 601 & 870
\end{tabular}
  \centering
  \caption{Parameters $d$ and $\Delta $ of the fits of the anomalous \tad\
  $\Delta \alpha$ of \lco\ (see Fig.~\ref{alp}) obtained for a LS/IS and for a LS/HS
   scenario with ($\nu$\, =\, 3)  and without ($\nu$\,=\, 1) orbital
   degeneracy of the excited IS (HS) state. The respective scaling factors $C$ of
   Eq.~\ref{scaling}
    are given in the last row. Experimentally we find $C^{exp}$\,=\,172\,emuK/mole.
    }\label{werte}
\end{table}

Within an ionic picture the LS/IS scenario in \lco\ is surprising. The energies
of the different spin states of Co$^{3+}$ are determined by the balance of the
Hund's rule coupling of parallel spins and the crystal-field splitting
$\Delta_{CF}$ between the \tg\ and the \eg\ levels. The ground state is either
the HS (small $\Delta_{CF}$) or the LS (large $\Delta_{CF}$) but never the IS
state. Moreover, the IS is expected to lie at least about 1\,eV above the
ground state~\cite{sugano}. In order to explain the much smaller value observed
experimentally ($\Delta$\,$ \simeq $\,$185$\,K\,$\simeq $\,$0.016$\,eV) the
energy of the IS state has to be lowered relative to the LS and HS states. This
can arise from a hybridization between the Co-\eg\ and the O-$2p$ levels as has
been found in LDA+U band-structure calculations~\cite{korotin96a}. We note that
an additional stabilization of the IS state can arise from a JT distortion. The
IS state is strongly JT-active because of the partially filled \eg\ level. The
HS state is only weakly JT-active since the gain of JT energy is much less in a
partially filled \tg\ level, and the LS is not JT-active at all. Thus, a JT
distortion favors the IS state and it easily explains the lifting of the
orbital degeneracy. Remarkably, this picture can also account for the behavior of
$\chi^C$ above 500~K. Due to the insulator-to-metal transition there are
delocalized charge carriers. The corresponding charge fluctuations usually
suppress (or weaken) the JT distortions. Then an orbital
degeneracy of $3$ has to appear above 500~K as suggested by our $\chi^C(T)$
analysis. In addition, an increase of the energy gap $\Delta$ may be expected
because of the loss of JT energy. However, the analysis of $\chi^C(T)$ above
600~K does not allow to determine a high-temperature value of $\Delta$
with sufficient accuracy. This latter point deserves further clarification.

In summary, our combined study of the magnetic \ch\ and the \tad\ shows that the
spin-state transition in \lco\ is consistently described by a thermal
population of the intermediate-spin state. The intermediate-spin state has no
orbital degeneracy up to about 500\,K. This may arise from (local) Jahn-Teller
distortions of the CoO$_6$ octahedra.
We analyzed our data within a simple ionic model of \lco\ but we stress that
the experimentally observed scaling between $\Delta \alpha $ and
$\partial (\chi \,T)\left/ \partial T  \right. $ is model-independent and may
serve as a sensitive test of more sophisticated models.

Valuable discussions with M.~Braden, A.~Freimuth,
M. Haverkort, Z.~Hu, E.~M\"{u}ller-Hartmann,
D. Khomskii, and L.H.~Tjeng are acknowledged.
This work was supported by the
Deutsche Forschungsgemeinschaft through SFB~608.


\begin{thebibliography}{10}
\bibitem{polarons}
Magnetization data of weakly Sr-doped \lco\ indicate the
formation of so-called high-spin polarons with S\,=\,10--16~\cite{yamaguchi96a}.
A similar effect can arise from oxygen non-stoichiometry. Thus one may roughly
estimate the impurity content $n$ by assuming a Curie-contribution of S\,=\,10
particles in which case $n$\,$<$0.1\%.
\bibitem{chargeorder}
The small $g$-factor could also be interpreted by a 'real'
  g-factor of the HS state of $\simeq$\,2 and the restriction that only
  25\,\% of the Co$^{3+}$ ions may thermally populate the HS state. This
  scenario would require a kind of 3\,LS~:~1\,HS state order,
  which is not favored by any of the experimental
  investigations on \lco .
\end{thebibliography}
\end{document}